# TEACHING RESULT ANALYSIS USING ROUGH SETS AND DATA MINING

[1]P.Ramasubramanian, [2]K.Iyakutti, [3]P.Thangavelu, [4]J.Joy Winston

[1]Professor, Dept. of CSE, Dr.G.U.Pope College of Engineering, Sawyerpuram – 628251. Thoothukudi, Tamilnadu – India.
[2]CSIR Emeritus Scientist, School of Physics, Madurai Kamaraj University, Madurai – 625025. Tamilnadu – India.
[3]Research Department of Mathematics, Aditanar College of Arts & Science, Tiruchendur – 628216.Tamilnadu–India.
[4]Lecturer, Dept. of CSE, Dr.G.U.Pope College of Engineering, Sawyerpuram – 628251. Thoothukudi, Tamilnadu – India.

**Abstract -** The development of IT and WWW provides different teaching strategies, which are chosen by teachers. Students can acquire knowledge through different learning models. The problem based learning is a popular teaching strategy for teachers. Based on the educational theory, students increase their learning motivation, which can increase learning effectiveness. In this paper, we propose a concept map for each student and staff. This map finds the result of the subjects and also recommends a sequence of remedial teaching. Here, rough set theory is used for dealing with uncertainty in the hidden pattern of data. For each competence the lower and upper approximations are calculated based on the brainstorm maps.

## 1. Literature Review

A teacher always wants to know how to evaluate the knowledge of the students in the teaching process. The teacher can understand the mental model of a student through brainstorm map, which emphasizes on group discussion and the teacher can ask the students to draw their own brainstorm maps about the problem [1]. Moreover, the teacher is a only facilitator to students in the teaching process[2].

This paper proposes a tool to analyze the results of the subject based on the teaching taught by the teacher. The brainstorm map is created when the students are trying to solve the problems proposed by the teacher in a class. Since the brainstorm map represents the concepts in student's minds concretely, teacher can use the brainstorm map to check what students have learned in class and can guide in the learning direction. Moreover, the teacher can adopt his/her teaching strategies according to the brainstorm map.

In a educational system, the student may belong to the group of categories. According to the category, the student's learning method is varied. The Teacher should be aware of their knowledge and the teacher must provide the teaching materials to them. For improving their results, we form an integrated map by combining both student and teacher maps. The concept map can help the teacher to know the weakness of students. Based on the result, the teacher can adjust his teaching strategies.

## 2. Introduction to Rough Set

The Rough Set Theory is a recent mathematical theory employed as a data mining tool with many favorable advantages. Since this theory has been applied to various domains, the majority of these applications are used to solve the classification problems, which exclude the temporal factor in data sets. The rough set analysis is presented as a technique to direct the knowledge discovery process from data.

In 1982, Pawlak introduced the Rough Set Theory(Z.Pawlak [7]). This rough set theory was initially developed for a finite universe of discourse in which the knowledge base is a partition, which is obtained by any equivalence relation defined on the universe of discourse and it is useful for knowledge discovery and reasoning about imprecise or incomplete data and to discover the hidden data pattern [3]. The basic idea of rough set is raw data and rough data. Raw data is often very detailed, disorganized, incomplete and imprecise. Rough data can be more efficient, effective and robust. Roughest methodology is a new technique for reasoning from imprecise and ambiguous data.

Let U be any finite universe of discourse. Let R be any equivalence relation defined on U. Clearly, the equivalence relation partitions U. The pair (U,R) is called the approximation space. This collection of equivalence classes of R is called as knowledge base. Then for any subset A of U, the lower and upper approximations are defined respectively as follows:

$$\underline{R}A = \cup \{W_i : W_i \subseteq A\}$$
$$\overline{R}A = \cup \{W_i : W_i \cap A \neq \varphi\}$$

The ordered pair $(\underline{R}A, \overline{R}A)$ is called rough set. In general, $\underline{R}A \subseteq A \subseteq \overline{R}A$. If $\underline{R}A = \overline{R}A$ then A is called exact. The lower approximation of A is called the positive region of A and is denoted by POS(A) and the complement of upper approximation of A is called the negative region of A and is denoted by NEG(A). Its boundary is defined as BND(A) = $\overline{R}A - \underline{R}A$. Hence, it is trivial that if BND(A)= $\varphi$, then A is exact.

In the theory of rough sets, the decision table of any information system is given by T=(U,A,C,D), where U is the universe of discourse, A is a set of primitive features, C and D are the subsets of A, called condition and decision features respectively.

For any subset P of A, a binary relation IND(P), called the indiscernibility relation is defined as IND(P) = {(x,y) $\in U \times U : a(x) = a(y)$ for all a in P}

The Rough set in A is the family of all subsets of U having the same lower and upper approximations. Usually rough set theory are explored using the decision tables i.e., information tables.

In this paper, the data in the concept map are divided into three parts namely upper approximation (UP), lower approximation (LOW) and boundary region (BND). For example, the data belongs to UP means, the data is consistent and the data belongs to LOW means the data is inconsistent with their expectation. The BND is a set of nodes that are parents of nodes in UP and LOW.

The UP and BND sets will be taken into consideration while developing a mechanism to find the sequence of the lacks of the student. In order to assist the teacher in problem based environment, some interaction must be retrieved from the concept maps of the students and the teacher. This must be discovered at first. For example, from the concept maps that a student might be missing could be found. However, finding the hidden information inside the concept map is difficult. The rough set theory is usually used to discover the hidden data pattern.

3. **Relationship between concept map and rough set**

The brainstorm map is very hard for teachers to check if the concepts in student's mind are either correct or partially correct, since the brainstorm map is too complex to understand. A concept map includes nodes (terms or concepts), linking lines and linking phrases which describe the relationship between nodes. Concept map always assumed a hierarchy. General and rough concepts are located in higher level [4].

In an integrated map, the green circle indicates that the concepts exist in the minds of student and teacher, the red circle representing the concept is missing in the student's mind. The element of UP and LOW sets in each level will be the green circle and the red circle. Regarding the boundary set, we have taken as parent node of the elements of the UP and LOW sets as the elements of the BND set. The following three conditions might help to construct the integrated concept map.

i. The concept is consistent in the teacher's map but not consistent in the student's map
ii. The concept is inconsistent in the teacher's map but consistent in the student's map.
iii. The concept map is consistent in the teacher's map and in the student's map,

Teacher will easily find the concepts in the integrated concept map with different colors. In condition (i), the concept node will be in red. In condition (ii), the concept node will be in green and in condition (iii) the concept node will be in green.

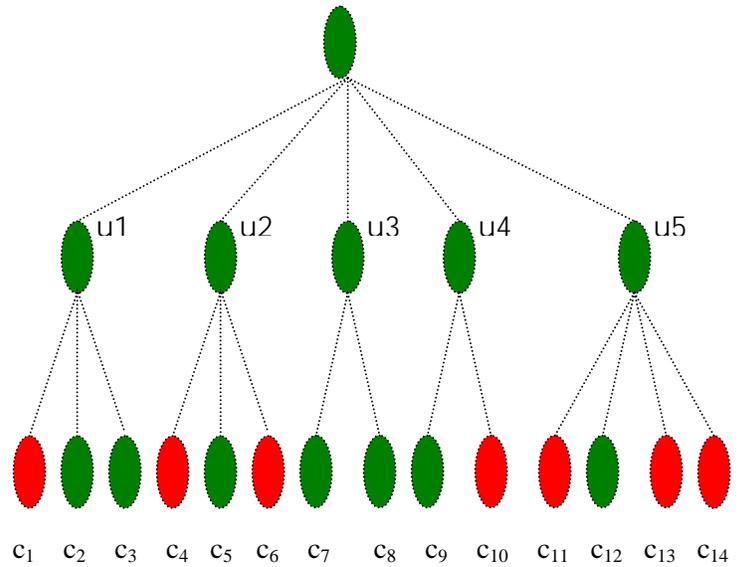

(ii) Concept map of Student

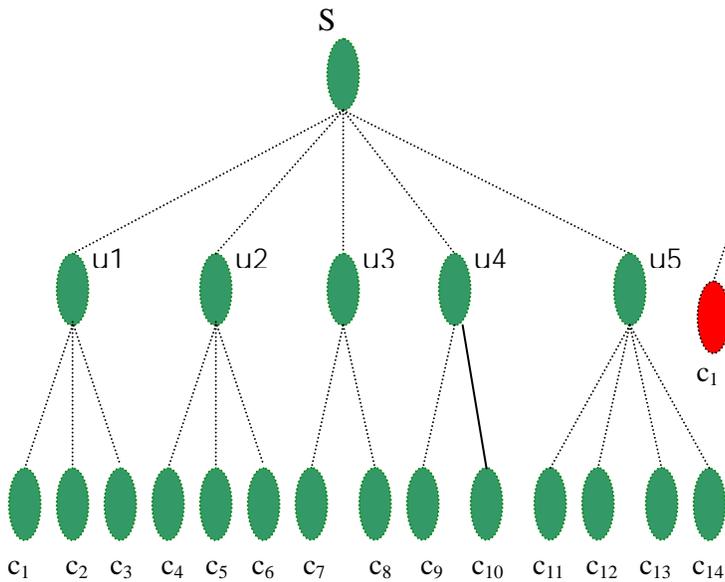

(i) Concept Map of Teacher

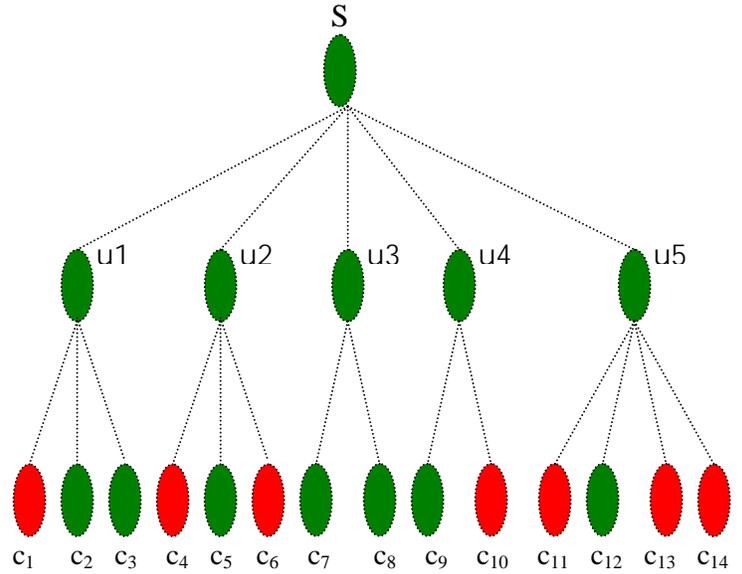

(ii) Integrated map of both student and teacher

**Figure 1: Concept Map**

The teacher will understand the lack of concept in students easily and provide the necessary supplemental materials to students according to the lack of concepts and also find the results. Each of the lack of concept will be assigned to a value with the sequencing algorithm. After an integrated map is constructed the rough set theory can be used to analyze the difference between student and teacher.

## 4. Algorithm

The algorithm proposed in this paper will calculate the bottom up important degree of each concept except the leave nodes in the integrated map. We design a formula to calculate the feedback mechanisms.

$$\alpha_{level}(C_i) = \frac{|Children(C_i) \cap POS_{level+1}|}{|Children(C_i)|}$$

Where,

BND ($C_i$) = Weight of the boundary set.

The concept set is defined as,
$\{C_i\}=\{S_1,U_1,U_2,U_3,U_4,U_5,C_1,C_2,C_3,C_4,C_5,C_6,C_7,C_8,C_9,C_{10},C_{11},C_{12},C_{13},C_{14}\}$

The children concept set of $C_i$ is Children ($C_i$)

Children($C_i$)=$\{S_1,U_1,U_2,U_3,U_4,U_5,C_2,C_3,C_5,C_7,C_8,C_9,C_{12}\}$

The upper approximation, Lower approximation and Boundary set of each level will be denoted as UP $_{level}$(C), LOW $_{level}$(C) and BND(C). The elements of UP, LOW and BND set for each level is shown below.

UP $_{level}$(C) = $\{C_2, C_3, C_5, C_7, C_8, C_9, C_{12}\}$
LOW $_{level}$(C) = $\{C_1, C_4, C_6, C_{10}, C_{11}, C_{13}, C_{14}\}$
BND(C) = $\{U_1, U_2, U_3, U_4, U_5\}$

## 5. Rough set and their elements

First in order to bypass the leave nodes of the integrated map, those leave nodes are identified. Since the deepest level of the integrated map is 2, the boundary set of level 2 is $BND_2(C)$ will be processed. This is shown in Table 1.

**Table 1**

| $POS_2(C)$ | C2, C3, C5, C6, C7, C8 |
|---|---|
| $NEG_2(C)$ | C1, C4, C9 |
| $BND_2(C)$ | U1, U2, U3, U4, U5 |
| $POS_1(C)$ | U1, U3, U4, U5 |
| $NEG_1(C)$ | U2 |
| $BND_1(C)$ | S1 |

Using the above table, the teacher can decide how much percentage of the results he/she can achieve and based on that he/she can also decide to provide the supplementary teaching materials in sequence to either a student or a group by the important degree of the concept. He/She can decide the order from either small to large (0.00 to 0.99) or large to small (0.99 to 0.00). If the important degree of concept node is 1.00, the whole branch from the concept is known clearly by the student. The result is calculated by the formula as shown in table 1.

**Table 2: Result Analysis for Student Feedback Mechanism**

| BND Set | Level | $|Child(C_i)|$ | $|Children(C_i) \cap POS_{level+1}|$ | $\alpha_{level}(C_i)$ |
|---|---|---|---|---|
| $U_1$ | 2 | 3 | 2 | 2/3=0.66 |
| $U_2$ | 2 | 3 | 1 | 1/3=0.33 |
| $U_3$ | 2 | 2 | 2 | 2/2=1 |
| $U_4$ | 2 | 2 | 1 | 1/2=0.5 |
| $U_5$ | 2 | 4 | 1 | 1/4=0.25 |

Expected Result = $\frac{\text{total}}{\text{size}} = \frac{2.74}{5} = 0.548$.

The actual result is compared with the expected result which is shown in table 2. The grade is provided to that result for separating the student and providing the teaching material according to their skill. If the result is greater than or equal to 75% then grade is 'A' or else if the result is 1 greater than or equal to 50 then grade is 'B', else if the result is less than 50%, then the grade is 'C'.

**Table 2 : Assigning Grade to Student Feedback Mechanism**

| BND SET | EXPECTED RESULT (In %) | ACTUAL RESULT (In %) | GRADE |
|---|---|---|---|
| $U_1$ | 100 | 66 | B |
| $U_2$ | 100 | 33 | C |
| $U_3$ | 100 | 100 | A |
| $U_4$ | 100 | 50 | B |
| $U_5$ | 100 | 25 | C |

## 6. Experiments and Discussions

Our experiment system focuses on providing supplemental teaching materials for teacher and analyzes the results. By using this technique based on the student feedback systems, a teacher can revise his/her teaching plan and may improve his/her results. We collected student details such as their Register number, name, Department, details of the subjects for corresponding semester for examining our feedback system and acquire some processing results. The experiment system is implemented in VB .Net 2003 based application environment.

In this study, the teacher is a facilitator and who can't make learning activities with students anytime. The teacher only provides to the students clues for finding and solving problems in most of the time. The teacher hopes to understand what major subject is discussed by the students and how they understand the subjects. In this paper, we provide a feedback system for the teacher to understand the students' learning situation. By using concept map technique, how much percentage the students understand the subject is known and the teacher can determine what is necessary for students. From that we can improve the result of the students.

## 7. Conclusion

In this paper, we analyzed the technique for concept map method of teaching. It makes teacher know what students are thinking and how to modify his/her teaching method. The major contribution of this paper is to help teacher find what is the lack in concepts of students and immediately the teacher can supply to them. It is very easy for him/her to realize the lack of concept in students and provide necessary assistance to the students. Using the technique discussed in this paper, one can access the performances in a particular subject which is accessed by a teacher.

5. Rosetta, Rough set toolkit at http://www.idi.ntnu.no/~aleks/ Rosetta/.
6. Ziarko, W. Variable precision rough set model, Journal of Computer and System Sciences 46: 39-59, 1993.
7. Z.Pawlak, "Rough Sets", International Journal of Computer and Information Sciences, Pages 341 – 356, 1982.

**About the Authors**

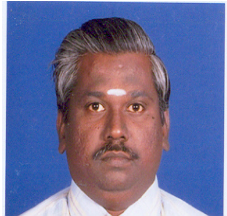

**P.Ramasubramanian**, is Professor in the Department of Computer Science and Engineering at Dr.G.U.Pope College of Engineering, Sawyerpuram, Thoothukudi, Tamilnadu, India. He obtained his Bachelor and Master degree in Computer Science and Engineering from Madurai Kamaraj University, Madurai in the year 1989 and 1996 respectively. He is doing Ph.D in Madurai Kamaraj University, Madurai. He has over 21 years of Teaching Experience and authored 15 books and 8 research papers in International and National Journals and 11 papers in International and National Conferences. His current area of research includes Data Mining, Data Ware housing, Neural Networks and Fuzzy logic. He is a life member of ISTE, Fellow in Institution of Engineers and members in other professional societies.

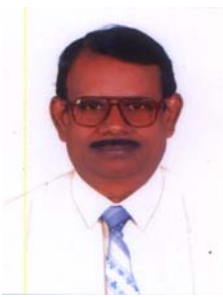

**Dr. K.Iyakutti, is CSIR Emeritus Scientist, School of Physics,** in the Madurai Kamaraj University. After getting Ph.D from **University of Madras**, he was a Lecturer there in the Department of Nuclear Physics. He joined as Reader in Madurai Kamaraj University in 1984 and was appointed as Professor in 1987. He has more than **30 years of Service in Teaching (PG) and Research.** He has guided many Ph.D and M.Phil Scholars. He is a recognized guide in computer Science in Madurai Kamaraj University. His area of Research is computational Physics including Nanotechnology. He has authored more than **125 research papers** in International, National Journals and Conferences. He has been **Principal and Co-investigator** in many research schemes funded by **CSIR, UGC, DAE and DST.** He was a **Post Doctoral Fellow** in the Department of Quantum Chemistry**, Uppsala University, Sweden** and he has visited many Research centers abroad including **ICTP, Trieste, Italy, Rensselaer Polytechnic Institue, Troy, USA and National Research Council of Canada.** He has attended many international and National Seminars, Workshops, Summer Schools and Conferences. He was responsible for the initiation of **Electronic Band structure Research** in South India. He is the associate editor of the Book in Tamil **"Encyclopedia of Physics"** edited by Prof. P.K.Ponnuswamy, published by University of Madras in 1977.

He was **Hon. Joint Director, Science Education Center**, Madurai Kamaraj University during 1986-1989. He was the **Head and Coordinator** of the School of Physics during 1995-97. He was the **Syndicate member** (elected) of the Madurai Kamaraj University, and member of the Finance committee.

**Dr. P.Thangavelu,** working as Reader in the Research Department of Mathematics, Aditanar 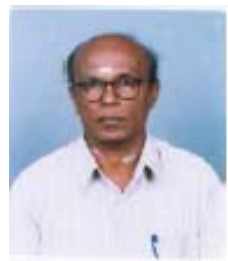 College of Arts and Science, affiliated to Manonmanian Sundaranar University, Tirunelveli, India. He got his Ph.D **degree from Alagappa University, Karaikudi, Tamilnadu, India.** He is teaching Mathematics and allied fields to UG, PG and M.Phil students and guiding Ph.D scholars in Mathematics. His current area of research includes Analysis, Topology, Fuzzy Topology, Neural Networks, Data Mining, Rough Set Theory, Fixed Point Theory and Artificial Intelligence. **He published more than 30 papers in mathematics** and **allied fields**. **He was the Chairman, PG Board of Studies in Mathematics in Manonmaniam Sundaranar University for the period April 2005 to March 2008.**

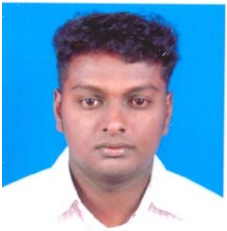

**J.Joy Winston**, is Lecturer in the Department of Computer Science and Engineering at Dr.G.U.Pope College of Engineering, Sawyerpuram, Thoothukudi, Tamilnadu, India. He obtained his Bachelor and Master degree in Computer Science and Engineering from Anna University Chennai and Tirunelveli in the year 2007 and 2009 respectively. He has 1 year Teaching Experience and published 1 paper in National Conference. His current area of research includes Data Mining, Data Ware housing, Neural Networks and Fuzzy logic. He is a life member of ISTE.